\documentclass[12pt,onecolumn,superscriptaddress,showpacs,preprintnumbers,amsmath,amssymb,prl,aps]{revtex4}
\usepackage{graphicx}

\begin{document}

\title{Thermodynamics and geometry of strange quark matter
}
\author{H. Gholizade}
\email{husen.qulizade@gmail.com}
\affiliation{Department of Physics, Tampere University of Technology
P.O.Box 692, FI-33101 Tampere, Finland.}

\author{A. Altaibayeva}
\email{aziza.ltaibayeva@gmail.com}
\affiliation{Eurasian International Center for Theoretical Physics and Department of General \&
Theoretical Physics, Eurasian National University,
Astana 010008, Kazakhstan}

\author{R. Myrzakulov}
\email{rmyrzakulov@gmail.com}
\affiliation{Eurasian International Center for Theoretical Physics and Department of General \&
Theoretical Physics, Eurasian National University,
Astana 010008, Kazakhstan}

\date{\today}

\begin{abstract}
We study thermodynamic of strange quark matter (SQM) using the analytic expressions of free  and internal energies. We investigate two regimes of the high density and low density separately. As a vital program, in the case of a massless gluon and massless quarks at finite temperature, we also present a geometry of thermodynamics for the gluon and Bosons  using a Legendre invariance metric, it is so called as geometrothermodynamic (GTD) to better understanding of the phase transition. The GTD metric and its second order scalar invariant have been obtained, and we clarify the phase transition by study the  singularities of the scalar curvature of this Riemannian metric. This method is ensemble dependence and to complete the phase transition. Meanwhile, we also investigate  enthalpy and  entropy and internal energy representations. Our work exposes new pictures of the nature of phase transitions in SQM.

\end{abstract}


\maketitle

\section{Introduction}

Quark matter idea introduced by Budmer \cite{budmer} and
Witten \cite{witten},  and the thermodynamics of quark matter
investigated by Jaff and Farhi \cite{jeff} and several authors
\cite{SQMref}. Only strange quark matter (SQM) - up,
 down and strange quarks- can be stable and non strange quark matter -up and down quarks- exists at an excited state respected to nuclear matter\cite{jeff}.
The quark matter can be formed in Heavy Ions Colliders (HIC), neutron stars and early universe \cite{5.6}.\par
 Quark matter can be in different phases due to values of  density and temperature of the system. At high temperature, the quark matter is in "Quark Gluon Plasma" (QGP) phase, without spontaneous symmetry breaking \cite{rev modern phys}. This phase is expected to exist in the early universe and also in HIC (small and short-lived)\cite{rev modern phys}.
If we denote by $T$ absolute temperature and $\mu$ chemical potential, at $T\ll \mu$ the spontaneous phase transition happens. We expected that this region exists in the neutron star's core. At high density and low-temperature region, we have quark liquid (degenerate Fermi gas). In this region, system can be in color superconductive phase and  ferromagnetic phase.
\cite{rev modern phys}\par
From another point of view, according to the equivalence principle of general relativity any matter field gravitates and
so SQM can be used as a source of gravity \cite{Harko:2013wka}.
It is one direction on how SQM gravitates. But in this work we are interested to describe the geometry of SQM in a simple case of the massless gluon using the same geometrical methods which are used to construct general relativity. Based on the Riemannian geometry we can construct metric for thermodynamic of SQM. We present a general formalism in the next sections. We use Legendre's invariant metric to describe the geometry of gluons. By study singularities of scalar curvatures, we investigate the possibility of interaction in gluons.  \par
Our plan in this paper is as the following: In Sec.2, we study thermodynamic of SQM. In Sec.3, Fermi gas at finite temperature in the relativistic regime will be studied. In Sec.4, We study massive Fermions in low-temperature regime $\mu>>T$. In Sec.5, basic properties of massless Fermions in regime $\frac{T}{m}\rightarrow\infty$ is calculated. In Sec.6,  Gluons at finite $T$ are investigated in details.  In Sec.7, is devoted to studying a geometric approach to the phase transitions in Legendre's  invariant formalism. We study the Riemannian geometry of Gluons, Bosons and Fermions. By a careful analysis of singularities, we find the critical points of these systems. We conclude in the final section.

\section{\label{sec:SQM}SQM thermodynamics}
In this section we present a short review of the thermodynamics of quark matter.
The thermodynamic potential density of massless quarks in second
the order is written as the following\cite{freedman}:
\begin{equation}
\Omega=-\frac{1}{4
\pi^2}\sum_{i=1}^{N_f}\mu_i^4[\frac{N_c}{3}-N_g\frac{\alpha_c}{\pi}].
\end{equation}\label{SQMeq1}
The ideal massive S quark thermodynamic potential is obtained as follows:
\begin{equation}
\Omega^s=-\frac{N_c}{3\pi^2}[\frac{1}{4}\mu_s(\mu_s^2-m_s^2)^{\frac{1}{2}}(\mu_s^2-\frac{5}{2}m_s^2)
+\frac{3}{8}m_s^4\ln[\frac{\mu_s+(\mu_s^2-m_s^2)^{\frac{1}{2}}}{m_s}]].
\end{equation}\label{SQMeq2}
The first order perturbative term $\alpha_c$  in thermodynamic potential has the meaning of  the exchange interaction. For massive quarks  exchange interaction
contribution in thermodynamic potential is written as the below:
\begin{equation}
\Omega_{ex}=\frac{N_g\alpha_c}{\pi^3}\left\{\frac{3}{4}\left[\mu_s(\mu_s^2-m_s^2)^{\frac{1}{2}}-m_s^2
\ln[\frac{\mu_s+(\mu_s^2-m_s^2)^{\frac{1}{2}}}{m_s}]\right]^2-\frac{1}{2}(\mu_s^2-m_s^2)^2\right\}.
\end{equation}\label{SQMeq3}

At zero temperature and free SQM we obtain following results for thermodynamics potentials density:
\begin{eqnarray}\label{SQM6}
\Omega_u=-\frac{\mu_u^2}{4\pi^2},\ \
\Omega_d=-\frac{\mu_d^2}{4\pi^2},\\ \nonumber
\Omega_s=-\frac{\mu_s^2}{4\pi^2}[(1-\lambda^2)^{1/2}(1-\frac{5}{2}\lambda^2)
+\frac{3}{2}\lambda^4\ln[\frac{1+(1-\lambda^2)^{1/2}}{\lambda}]].
\end{eqnarray}
$\lambda=\frac{m_s}{\mu_s}$ and $m$ is mass of  strange quarks.
After some algebraic manipulations, we obtain the total internal energy s:
\begin{eqnarray}\label{SQM10}
U(N_b,V)= B V + \frac{5 m^2 V\sqrt {4\pi^4 - \frac{9 m^2 V^2}{
{N_b}^2}}}{16\pi^4} - \frac{{N_b}^2\left(-8\pi^2 + \sqrt {4\pi^4 -
\frac{9 m^2 V^2}{{N_b}^2}} \right)}{18 V}\\ \nonumber - \frac{27 m^4
V^3\ln\left[-\frac{2\pi^2}{-2\pi^2 + \sqrt{4\pi^4 -\frac{9 m^2
V^2}{{N_b}^2}}} \right]}{32{N_b}^2\pi^6}.
\end{eqnarray}
Finite temperature calculations for massless quarks are simple, but for massive quarks we can use the low temperature results. At finite temperature we must add the gluons effects in thermodynamics, because their thermodynamic potentials at finite temperature are not zero. All results of this section are motivated due to the leakage of analytical expressions in literature.

\section{\label{subsec 1} Massive Fermions At Low Temperature}
The particle number density of the relativistic Fermi system is:
\begin{equation} \label{teq7}
n =\frac{g}{2\pi^2}
 \int_0^{\infty}
 \left[
   \frac{1}{1+e^{(\epsilon-\mu)/T}}
  -\frac{1}{1+e^{(\epsilon+\mu)/T}}
 \right]
p^2 \mbox{d}p.
\end{equation}
The second term in brackets is anti-Fermions number density. At the low temperature limit ($\mu \gg T$) its contribution to the Fermions number density is negligible, so we can write following relations for thermodynamics quantities \cite{royalsoc}:
\begin{eqnarray}\label{teq8}
N=\frac{g V}{2 \pi^2}\int_0^{\infty}\frac{(x^2+2 m^2)^{\frac{1}{2}}(x+m)}{\exp[\frac{x-\mu}{T}]+1}dx\\
U=\frac{g V}{2 \pi^2}\int_0^{\infty}\frac{x(x^2+2 m^2)^{\frac{1}{2}}(x+m)}{\exp[\frac{x-\mu}{T}]+1}dx\\
p=\frac{g V}{6 \pi^2}\int_0^{\infty}\frac{(x^2+2 m^2)^{\frac{3}{2}}}{\exp[\frac{x-\mu}{T}]+1}dx.
\end{eqnarray}
Expanding the integrals at the low temperature limit we have \cite{royalsoc}:
\begin{eqnarray}
\mu=\mu_0\{1-\frac{\pi^2}{6}(\frac{T}{\mu_0})^2\frac{(1+2 \xi
^2)[(1+\xi^2)^{\frac{1}{2}}-1]}{\xi^2(1+\xi^2)^{\frac{1}{2}}}
-\frac{\pi^4}{360}(\frac{T}{\mu_0})^4\frac{(5\xi^2+4)
(4\xi^2+9)[(1+\xi^2)^{\frac{1}{2}}-1]^3}{\xi^6(1+\xi^2)^{\frac{3}{2}}}\},\\
\nonumber \xi=\frac{2\pi}{m}(\frac{3 n}{4\pi g})^{\frac{1}{3}},\ \
\mu_0=m[(1+\xi^2)^{\frac{1}{2}}-1].
\end{eqnarray}
Using the similar low temperature expansion method we can write following relations for other thermodynamic potentials \cite{royalsoc}:
\begin{eqnarray}
\label{teq10}
U=3 N m \left [ \varepsilon(\xi)+2
\frac{\pi^2}{12}(\frac{T}{m})^2\frac{(1+\xi^2)^{\frac{1}{2}}}{\xi^2}+
6
(\frac{T}{m})^4\frac{(1+\xi^2)^{\frac{1}{2}}}{\xi^6}\left\{\frac{7\pi^4}{720}(2\xi^2-1)-\frac{\pi^4}
{144}\frac{(2\xi^2+1)^2}{1+\xi^2}\right\} \right ]
\end{eqnarray}
and the heat capacity at low temperature is: \cite{royalsoc}:
\begin{eqnarray}
C_v=&&\frac{\partial U}{\partial T}|_v
=12 N\frac{T}{m}\left[
\frac{\pi^2(1+\xi^2)^{\frac{1}{2}}}{\xi^2}+6 (\frac{T}{m})^2
\frac{(1+\xi^2)^{\frac{1}{2}}}{\xi^6}
\left\{\frac{7\pi^4}{720}(2\xi^2-1)
-\frac{\pi^4}{144}\frac{(2\xi^2+1)^2}{1+\xi^2}\right\}\right].
\end{eqnarray}

For massive Fermions up to order two for isothermal compressibility, we have:
\begin{eqnarray}
\kappa=\frac{54 g m V^2 \left(1+6^{2/3} \pi ^{4/3} \left(\frac{N}{g V}\right)^{2/3}\right)^{3/2} \sqrt{1+\frac{6^{2/3} \pi ^{4/3} \left(\frac{N}{g V}\right)^{2/3}}{m^2}}}{\mathcal{A}}
\end{eqnarray}
Where
\begin{eqnarray}
\mathcal{A}=108\ 6^{1/3} \pi ^{8/3} N^2 \left(\frac{N}{g V}\right)^{1/3} \sqrt{1+6^{2/3} \pi ^{4/3} \left(\frac{N}{g V}\right)^{2/3}}\\ \nonumber
+18\ 6^{2/3} g \pi ^{4/3} V N \left(\frac{N}{g V}\right)^{2/3} \sqrt{1+6^{2/3} \pi ^{4/3} \left(\frac{N}{g V}\right)^{2/3}}
+9 g \pi ^2 T^2 V N \sqrt{1+\frac{6^{2/3} \pi ^{4/3} \left(\frac{N}{g V}\right)^{2/3}}{m^2}}\\  \nonumber
+\frac{6^{1/3} g \pi ^{2/3} T^2 V N \sqrt{1+\frac{6^{2/3} \pi ^{4/3} \left(\frac{N}{g V}\right)^{2/3}}{m^2}}}{\left(\frac{N}{g V}\right)^{2/3}}
+6\ 6^{2/3} g \pi ^{10/3} T^2 V N \left(\frac{N}{g V}\right)^{2/3} \sqrt{1+\frac{6^{2/3} \pi ^{4/3} \left(\frac{N}{g V}\right)^{2/3}}{m^2}}.
\end{eqnarray}
It is obvious that in zero temperature limit $T=0$, we have
\begin{eqnarray}
\kappa_{T=0}=\frac{3^{1/3} m \sqrt{1+\frac{6^{2/3} \pi ^{4/3} \left(\frac{N}{g V}\right)^{2/3}}{m^2}}}{2^{2/3} g \pi ^{4/3} \left(\frac{N}{g V}\right)^{5/3}}
\end{eqnarray}

\section{\label{subsec 2}Massless Fermions At Finite Temperature}
By following the method was proposed in \cite{astrophys,landau}, the internal energy of massless Fermions is:
\begin{eqnarray}\label{teq14}
U=3 p V = \frac{g
V}{8\pi^2}\left[\mu^4+2\pi^2\mu^2T^2+\frac{7\pi^4}{15}T^4\right].
\end{eqnarray}
The derivative of internal energy with respect to the
temperature at constant volume is written as the following:
\begin{eqnarray}\label{teq18}
C_v(T,V,\mu)=\frac{g V \left(\frac{28 \pi ^4 T^3}{15}+4 \pi ^2 T \mu
^2+4 \pi ^2 T^2 \mu  \mu '+4 \mu ^3 \mu '\right)}{8 \pi ^2},\\
\nonumber
\mu'=\frac{\partial \mu}{\partial T}=\frac{3^{1/3} g^5 \pi
^{8/3} T^5 V^5}{\sqrt{729 g^4 N^2 V^4+3 g^6 \pi ^2 T^6 V^6} \left(27
g^2 N
V^2+\sqrt{729 g^4 N^2 V^4+3 g^6 \pi ^2 T^6 V^6}\right)^{2/3}}\\
\nonumber +\frac{9 g^7 \pi ^{10/3} T^7 V^7}{\sqrt{729 g^4 N^2 V^4+3
g^6 \pi ^2 T^6 V^6} \left(81 g^2 N V^2+3 \sqrt{729 g^4 N^2 V^4+3 g^6
\pi ^2 T^6 V^6}\right)^{4/3}}\\ \nonumber -\frac{2 g \pi ^{4/3} T
V}{\left(81 g^2 N V^2+3 \sqrt{729 g^4 N^2 V^4+3 g^6 \pi ^2 T^6
V^6}\right)^{1/3}}.
\end{eqnarray}
The results for massless Fermions at finite temperature are valid for all temperature ranges, but in the next section , we shall calculate massive Fermions at low temperature. In that regime, the equations only valid at the low temperature limit and in this region we shall ignore the anti-Fermions contributions in  the equation of state.
\section{\label{subsec 3}Massless Gluons At Finite Temperature}
At finite temperature we must consider the gluons
contribution in thermodynamics parameters. In this section we ignore
the gluons condensation effects and higher order corrections. The
thermodynamic potentials of massless Bosons at finite temperature
are \cite{freedman}:
\begin{eqnarray}\label{teq19}
\Omega(T,V)=-\frac{8}{45} \pi ^2 T^4 V.
\end{eqnarray}
Heat capacity at finite temperature can written as:
\begin{equation}\label{teq20}
C_v^{gluon}(T,V)=\frac{32 \pi^2}{15} T^3 V.
\end{equation}
Collecting all results we can write the equation of state of free SQM at the low temperature limit.
Because the thermodynamic potentials are extensive thermodynamic quantities, we can add them in different phases. Therefore the total grand thermodynamic potential of SQM that composed of massive strange quarks and massless up, down quarks and gluons at the low temperature limit is:
\begin{eqnarray}\label{teq22}
\Omega_{SQM}=\sum_{i}^3\Omega_i+\Omega_{gluon} + B V,\\ \nonumber
C_v^{SQM}=\sum_{i}^3C_v^i+C_v^{gluon}.
\end{eqnarray}
For SQM we have $g=3 \times 2 =6$.
Again, like zero temperature case, we can calculate the interaction contributions in the equation of state, but the effects of all perturbation corrections can collected into an effective bag pressure \cite{7.10}.\par
Isothermal compressibility for  massless gluons by definition is:
\begin{equation}
\kappa=0.
\end{equation}
The thermodynamic relation between heat capacities is written as the following:
\begin{eqnarray*}
C_p=C_v+ T V \frac{\alpha^2}{\kappa}.
\end{eqnarray*}
About isobaric expansion coefficient we mention here that since the pressure depends only  on temperature, at constant pressure, temperature must remain constant, so the heat capacity at finite pressure ($C_p$) is meaningless. For free massless gluons constant pressure and constant temperature are equivalent.\par
The Isothermal expansion coefficient for massless gluons are also zero, but in the limit of $T V \frac{\alpha^2}{\kappa}$ it is important for us to discuss about $C_p$.  The constant heat capacity for gluons can be evaluated if we insert the interactions in the equation of state with medium dependent coupling constant. So  we ignore discussion about $C_p$, and only investigate other response functions.
\par
All the thermodynamics potentials are extensive parameters and we can add them in various phases. But isothermal expansion coefficients and compressibility can not be added in different phases (intensive parameters). All differentiations are with respect to  pressure or temperature of the system.
The volume of the system is: $\frac{\partial H^{SQM}}{\partial P_{SQM}}$, where enthalpy and pressure of SQM are the sum of enthalpy and pressure of all phases (massive and massless Fermions and massless gluons). So we can not add expansion coefficients in separate phases to each other.\par
 Unlike isothermal expansion coefficients and compressibility, the heat capacity at constant volume is an extensive parameter and we can simply calculate it for SQM.

\section{Legendre invariant Riemannian geometry  of massless Gluons and Bosons at finite temperature}
The relation
between  thermodynamic and gravitation is discussed in detail in several works, especially in the relation of the recent observational data in cosmology \cite{Bamba:2012cp}. There are two types of problems can be solved using differential geometry: one is how to describe matter fields in the presence of gravitational field using statistical mechanics. In this case you need to find the Green's functions of matter fields in curved space-time. Second is how to describe phase transitions in particle physics using geometry. In this sense, we need to find a unique metric associated with a set of thermodynamic variables and study the dynamics of this metric using Riemannian geometry. This type of problem has been investigated in this paper for quark matter. We use a Legendre invariant metric of the thermodynamic space to address the phase transitions. Let us start with some preliminary definitions:

\par
We describe a thermodynamic system with $n$ degree of the freedoms in an equilibrium state by a set of n extensive variables
$E^a$, with $a=1,...,n$. It means we have an equilibrium manifold and in any point of it , we can label a coordinate system. The fundamental equation which is needed is an energy function
 $\Phi=\Phi(E^a)$, here
$\Phi$ has the meaning of the thermodynamic potential \cite{callen}. We have different possibilities to identify this energy function for example
 the entropy $S$ or  the internal energy $U$
of the system in equilibrium. The differential geometric structure is not a unique and we can write several kinds of these forms. One of the first and simple and historically former is   a  Hessian metric as the following:
\begin{eqnarray}\label{Hessian}
g^{H} = \frac{\partial^2 \Phi}{\partial E^a \partial E^b } d E^a d E ^b \ ,
\end{eqnarray}
With this metric the manifold becomes
 Riemannian. It was presented in different ensemble forms previously \cite{old} . The Gauss second order form  $g^{H}$ looks like as a scalar under diffeomorphism  of coordinates $E^a \rightarrow \tilde E  ^a = \tilde E ^a (E^a)$.  It is possible to  change the potential using
the unique Legendre transformation \cite{arnold} :
\begin{eqnarray}
\{\Phi, E^a, I^a\} \longrightarrow \{\tilde \Phi, \tilde E ^a, \tilde I ^ a\}\ ,
\end{eqnarray}
\begin{eqnarray}
 \Phi = \tilde \Phi - \delta_{kl} \tilde E ^k \tilde I ^l \ ,\quad
 E^i = - \tilde I ^ {i}, \ \
E^j = \tilde E ^j,\quad
 I^{i} = \tilde E ^ i , \ \
 I^j = \tilde I ^j \ ,
 \label{leg}
\end{eqnarray}
where $I^a$ denotes a  set of $n$  intensive variables. It is known  the Hessian metrics are not invariant under Legendre transformations. The most general  Legendre's invariant  metric
can be written as the following:
\begin{equation}
G = \left(d\Phi - I_a dE^a\right)^2  +\Lambda
\left(\xi_{ab}E^{a}I^{b}\right)\left(\chi_{cd}dE^{c}dI^{d}\right) \  ,
\label{gup1}
\end{equation}
Here  $\xi_{ab}$ and $\chi_{ab}$ are diagonal constant
tensors, and $\Lambda$ denotes an arbitrary Legendre invariant function. We choice the metric to be in this form :
\begin{equation}
G^{II} =(d\Phi-\delta_{ab}I^{a}dE^{b})^{2}+ (\delta_{ab} E^{a}I^{b} )
(\eta_{cd}dE^{c}dI^{d})\ ,
\end{equation}
By computing the  pullback we have:
\begin{equation}\label{gII}
g^{II}= \left(E^a \frac{\partial \Phi}{\partial E^a}\right) \left( \eta_b^{ c} \, \frac{\partial^2\Phi}{\partial E^c \partial E^d} \, d E^b d E^d \right)
\end{equation}
It is the metric for the equilibrium manifold, here $\eta_b^{ c}={\rm diag} (-1,1,...,1)$. This formalism is called as the  geometrothermodynamics (GTD) \cite{quev07}.
 We need the fundamental equation $\Phi=\Phi(E^a)$. We believe the metric given by (\ref{gII}) is a useful tool to study the phase transitions in thermodynamical systems and the black holes and using this approach most of the critical points and even those ones that cannot be obtained by Davis's approach \cite{Davies} can be obtained as well \cite{applications}.\par
Metric for Bosons in $S$ representation with $E^a=\{U,V\},\gamma=\frac{3}{8} 5^{1/3} \left(\frac{3}{2 \pi}\right)^{2/3}$ is written:
\begin{eqnarray}
g^{II}_S\,={\frac {3}{16{\gamma}^{3/2}\sqrt {U{V}^{3}}}}\Big({V}^{2}dU^2- {U}^{2}dV^2\Big)
\end{eqnarray}
 Here curvature scalar and Riemannian tensor read:
\begin{eqnarray}
R^{II}_S=0,\ \ R^{IJKL}_S=0
\end{eqnarray}
So the geometry is flat and it indicates on no interaction. In fact the metric $g^{II}_S$ is conformally flat if we rewrite it in the following coordinates:
\begin{eqnarray}
u=\log U,\ \ v=\log V.
\end{eqnarray}
So, metric transforms to form:
\begin{eqnarray}
g^{II}_S\,=\frac {3}{16{\gamma}^{3/2}}e^{(v+3u)/2}(du^2-dv^2).
\end{eqnarray}
This metric is trvially conformal to another two dimensional flat metric $\eta=du^2-dv^2$. Vanishing of the curvature is also a result of this conformal flat duality.

\par
Similarly for $U$ representation in which $E^a=\{S,V\}$:

\begin{eqnarray}
g^{II}_U=\frac{4{\gamma}^{2} }{9}\left( {\frac {S}{V}} \right) ^{2/3}\Big(\frac{{S}^{2}dV^2}{V^2}-dS^2\Big).
\end{eqnarray}
By following the same technique as $S$ representation, we can show that this metric is also conformally flat under these coordinate transformations:
\begin{eqnarray}
s=\log S,\ \  v=\log V.
\end{eqnarray}
So, metric transforms to form:
\begin{eqnarray}
g^{II}_S=\frac{4{\gamma}^{2} }{9} e^{(8s-2v)/3}(-ds^2+dv^2).
\end{eqnarray}
 Here curvature scalar
\begin{eqnarray}
R^{II}_U=0.
\end{eqnarray}
So there exists no singularity in curvature. In fact, it is useful to check whether the metric is flat or not. So we compute Riemannian tensor $R_{IJKL}$, to find:
\begin{eqnarray}
R^{IJKL}_U=0.
\end{eqnarray}

\begin{figure*}
\includegraphics[width=15cm]{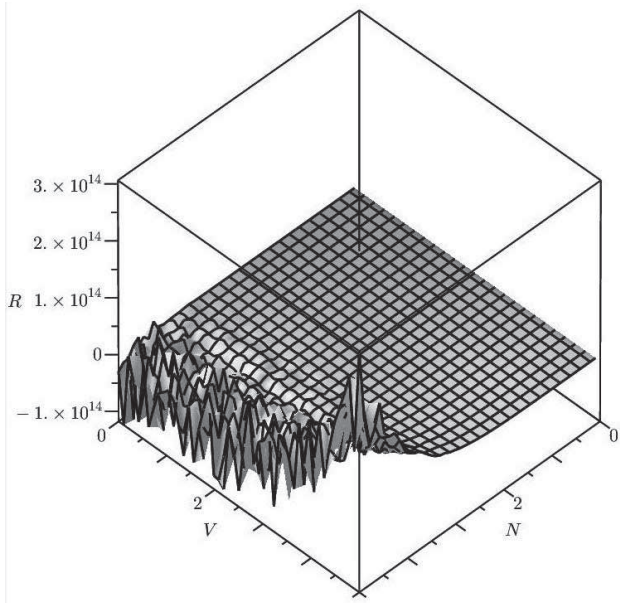}
\caption{ Ricci scalar for (\ref{SQM10}). It shows that for massless quarks the geometry is non trivial. We have singularities in R. So, quark matter enters the phase transition smoothly. It is one of the important evidences of GTD for quark matter.  }
\end{figure*}

\begin{figure*}
\includegraphics[width=15cm]{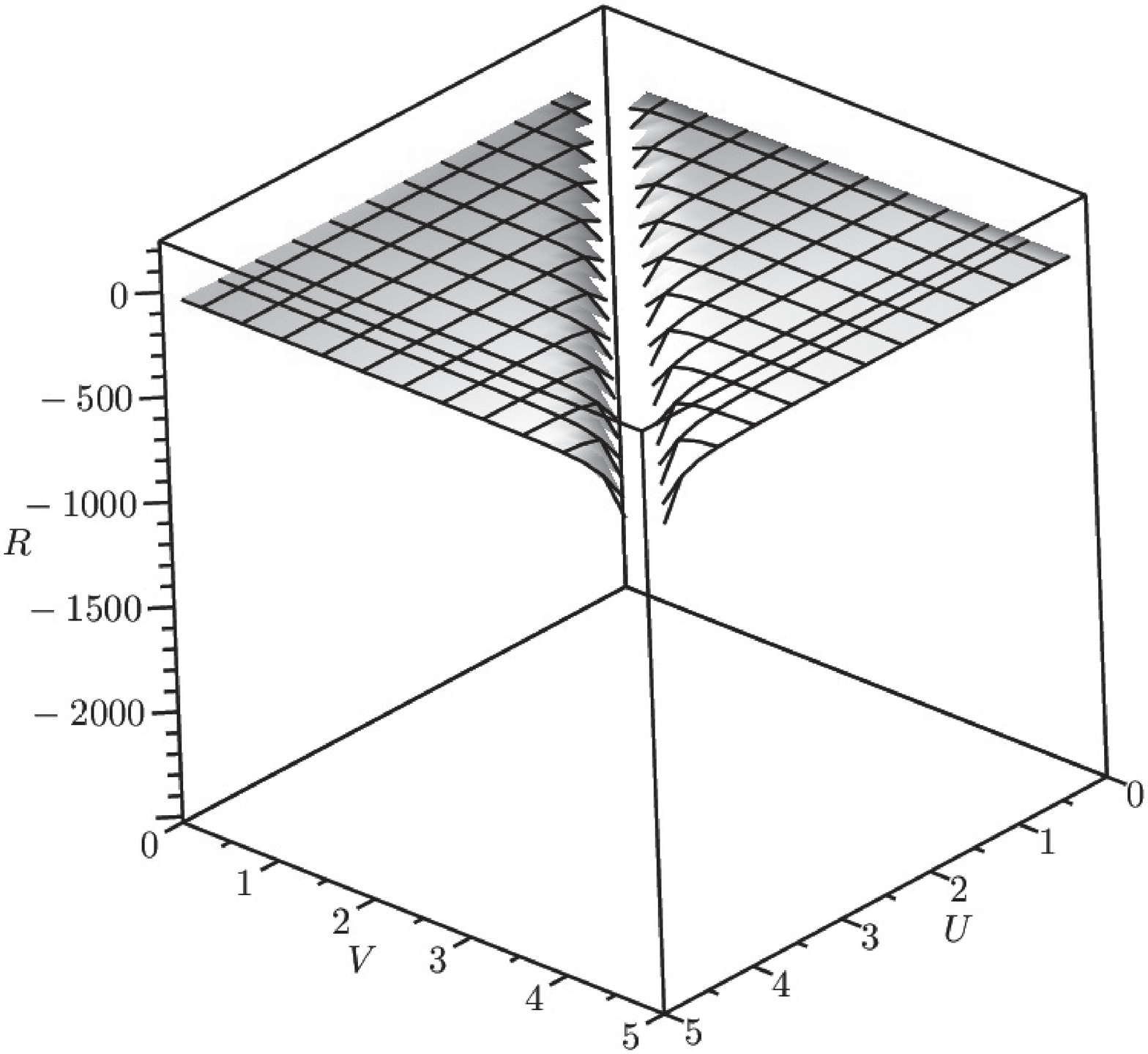}
\caption{ Ricci scalar for $R^{H}_S$. It shows that for massless quarks the geometry is non trivial. We have singularities in R.}
\end{figure*}

Consequently the geometry of Boson fields defined a flat manifold. So, according to the  GTD interpretation of interaction as geometrical curvature, flatness of manifold indicates to the no interaction system. So in the case of the massless Bosons the system has no interaction as GTD metric also predicts.\par
Just to show that GTD analysis depends to different ensembles (or not) we study GTD metric for $H(S,p)$, so we have
\begin{eqnarray}
g^{II}_{H-SS}\,=0.
\end{eqnarray}
So metric degenerates and we cannot apply GTD metric $g^{II}$ here.\par
It is adequate to deal with the case of Hessian metric given by (\ref{Hessian}). We have the following expressions for $g^{H}$ :
\begin{eqnarray}
g_{S}^{H}\,=\frac{3}{16{\gamma}^{3/4}} \left[ \begin {array}{cc} -\,{\frac {{U}^{3/4}}{{V}
^{7/4}}}&\,{\frac {1}{\sqrt [4]{U}{V}^{3/4}}}
\\ \noalign{\medskip}\,{\frac {1}{\sqrt [4]{U}{V}^{3
/4}}}&-\,{\frac {{U}^{3/4}}{{V}^{7/4}}}\end {array}
 \right]
     ,\
g_{U}^{H}\,=\frac{4\gamma}{9}\left[ \begin {array}{cc} \,{\frac {\,{S}^{4/3}}{{V}^{7/3}}}
&-\,{\frac {\,\sqrt [3]{S}}{{V}^{4/3}}}\\ \noalign{\medskip}-
\,{\frac {\,\sqrt [3]{S}}{{V}^{4/3}}}&\,{\frac {1}{
\sqrt [3]{V}{S}^{2/3}}}\end {array} \right]
.
\end{eqnarray}
So the curvature $R$ is written as the following:

\begin{eqnarray}
R^{H}_S= -\frac{4}{3}\,{\frac {{\gamma}^{3/4}{U}^{5/4} \left( -15\,{V}^{2}+7\,{U}^{2}
 \right) }{\sqrt [4]{V} \left( {U}^{2}-{V}^{2} \right) ^{2}}}
    ,\\R^{H}_U= \text{undefined}    .
\end{eqnarray}
Thus, curvature singularities exist only for $R^{H}_S$ and $g_{U}^{H}$ is singular. Singularities of  $R^{H}_S$ locate at $U=V$. FIG.2 shows a normalized behavior of $R^{H}_S$.
\par
Just to keep the motivation we plot the $R$ for (\ref{SQM10})  in FIG.2. This case defines GTD structure of massless quark. The system thermodynamically has phase transition. Geometry is non flat. Full expression of metric is so complicated and we shall not write them here. Using GR-TENSOR package for metric, we can obtain Ricci scalar R. We plot it in FIG.1. The Ricci scalar has singularities near some critical points $(N^*,V^*)$. Since GTD metric is not flat , we have interaction in the system. Further information has come from singularities in which $R$ diverges. GTD tool here predicts phase transition in the system as well as existence of interaction.

\section{Conclusion}
In this work we studied briefly basic thermodynamic properties of quark matters. In case of massless and massive we presented general expressions of energy functions and response functions. Our results are valid in both zero temperature and finite temperature limits. The Gluons and Bosons thermodynamics have been studied analytically. Further for the first time in literatures, we study geometry of some quark matter cases using a Legendre invariant formalism. For massless Gluons and Bosons and quarks in zero temperature, we study geometric metric. For Gluons and Bosons the geometry is conformal flat and we have not any type of interaction. But for the case of zero temperature, massless quark geometry has a more complicated non flat form. We computed Ricci scalar as the first scalar invariant of the associated metric. It showed that there exists interaction and also finite numbers of physical singularities in which $R $ diverges. This divergence predicts that system enters phase transitions.

\end{document}